\documentclass[12pt]{article}
\usepackage{latexsym,amssymb,amsmath}

\begin{document}

\title
{Remarks on string solitons}
\author
{E.K. Loginov\thanks{{\it E-mail:} ek.loginov@mail.ru.
Research was supported by RFBR Grant 06-02-16140.}\\
\it Department of Physics, Ivanovo State University\\
\it Ermaka St. 39, Ivanovo, 153025, Russia}
\date{}
\maketitle

\begin{abstract}
We consider generalized self-duality equations for $U(2r)$ Yang-Mills theory on $\mathbb R^8$
with quaternionic structure. We employ the extended ADHM method in eight dimensions to
construct exact soliton solutions of the low-energy effective theory of the heterotic string.
\end{abstract}

\section{Introduction}

In [1], an exact multi-fivebrane soliton solution of the heterotic string theory was
presented. This solution represented an exact extension of the three-level supersymmetric
fivebrane solutions of~[2]. Exactness is shown for the heterotic solution based on algebraic
effective action arguments and (4,4) worldsheet supersymmetry. The gauge sector of the
heterotic solution possesses $SU(2)$ instanton structure in the four-dimensional space
transverse to the fivebrane. An exact solution with $SU(2)\times SU(2)$ instanton structure
was found in~[3]. This soliton preserves four of the sixteen supersymmetries. In [4] a
one-brane solution of heterotic theory was found which is an everywhere smooth solution of the
equations of motion. The construction of this solution involves crucially the properties of
octonions. One of the many bizarre features of this soliton is that it preserves only one of
the sixteen space-time supersymmetries, in contrast to previously known examples of
supersymmetric solitons which all preserve half of the supersymmetries. A two-brane solution
of heterotic theory was found in [5,6]. This soliton preserves two of the sixteen
supersymmetries and hence corresponds to $N=1$ space-time supersymmetry in $(2+1)$ dimensions
transverse to the seven dimensions where the Yang-Mills instanton is defined. Some
generalization of one and two-brane solution was found in~[7,8]. All these solutions are
conformal to a flat space. In dimension six, the possibility of the existence of a
non-conformally flat solution on the complex Iwasawa manifold was discussed in~[9-11].
\par
In all above-named papers instanton solutions in various dimensions are extended to heterotic
string solitons. In this paper we employ the extended ADHM method in eight dimensions to
construct exact soliton solutions of the low-energy effective theory of the heterotic string.

\section{Generalized self-duality on $\mathbb R^{8}$}

In the first place, we define a basis $V_{\mu}$ with $(\mu)=(\mu_0,\mu_1)$ on $\mathbb
R^8\simeq\mathbb H\oplus\mathbb H$ as a collection of two quaternionic column vectors realized
as $4\times2$ matrices
\begin{equation}
V_{\mu_0}=\begin{pmatrix} e^{\dagger}_{\mu_0}\\0_2
\end{pmatrix}\qquad\text{and}\qquad
V_{\mu_1}=\begin{pmatrix} 0_2\\e^{\dagger}_{\mu_1}
\end{pmatrix},
\end{equation}
where $\mu_{k}$ is a four-valued index and the matrices
$(e^{\dagger}_{\mu_{k}})=(i\sigma_1,i\sigma_2,i\sigma_3,1)$. As in [12] we introduce the
anti-Hermitian matrices
\begin{equation}
N_{\mu\nu}=\frac12(V_{\mu}V_{\nu}^{\dagger}-V_{\nu}V_{\mu}^{\dagger}).
\end{equation}
Notice that for any $\mu,\nu=1,\dots,8$, we have $N_{\mu\nu}\in sp(2)$. To introduce
generalized self-duality equations on $\mathbb R^8$ we define the total antisymmetric tensor
\begin{equation}
T_{\mu\nu\rho\sigma}
=\frac{1}{12}\text{tr}(V_{\mu}^{\dagger}V_{[\nu}V_{\rho}^{\dagger}V_{\sigma]}).
\end{equation}
Then by direct calculation one finds that the matrix-valued tensor $N_{\mu\nu}$ is self-dual
in a sense of [13] (see also~[14]), i.e. it satisfies the eigenvalue equations
\begin{equation}
\frac12T_{\mu\nu\rho\sigma}N_{\rho\sigma}=N_{\mu\nu}.
\end{equation}
It is well known that the subgroup of $SO(8)$ which preserve the quaternionic structure and
therefore (4) is isomorphic to $Sp(1)\times Sp(2)/\mathbb Z_2$.
\par
With the help of the tensor (3) one may introduce an analog of the self-dual Yang-Mills
equations for $U(2r)$ gauge fields on $\mathbb R^8$. Indeed, if $F_{\mu\nu}$ is the
$su(2r)$-valued Yang-Mills field, then the generalized self-dual Yang-Mills equations in eight
dimensions is
\begin{equation}
\frac12T_{\mu\nu\rho\sigma}F_{\rho\sigma}=F_{\mu\nu}.
\end{equation}
Obviously, the equations (5) are invariant under $Sp(1)\times Sp(2)/\mathbb Z_2\subset SO(8)$
and any gauge field fulfilling (5) satisfies the second-order Yang-Mills equations due to the
Bianchi identities. In four dimention $T_{\mu\nu\rho\sigma}$ reduces to
$\varepsilon_{\mu\nu\rho\sigma}$ and, hence, (5) coincide with the standard self-dual
Yang-Mills equations.

\section{'t Hooft-type solutions in eight dimensions}

Now we construct a solution of the equations (5) (cf.~[15]). In the notations of the appendix
we choose $n=r=1$ and $k=2$. For the ADHM ingredients $a,b_{i}$ and $\Psi$ we propose the
ansatz
\begin{equation}
a=\begin{pmatrix} \Lambda 1_2 \\0_2
\end{pmatrix},\qquad
b_{i}=\begin{pmatrix} 0_2 \\1_2
\end{pmatrix}\qquad\text{and}\qquad
\Psi=\begin{pmatrix} \Psi_0\\\Psi_1
\end{pmatrix},
\end{equation}
where $\Lambda$ is a real constant and $i=0,1$. With selection we obtain
\begin{equation}
\Delta^{\dagger}\Delta=(\Lambda^2+x^{\dagger}x)\otimes 1_2,
\end{equation}
where $x=x_1+x_2=x^{\mu_{i}}e^{\dagger}_{\mu_{i}}$. It is obvious that the conditions (A.2)
and (A.3) are satisfied. Next, the equations (A.4) becomes
\begin{equation}
\Lambda\Psi_0+x^{\dagger}\Psi_1=0_2,
\end{equation}
which is solved by the solution
\begin{equation}
\Psi_0=\varphi^{-1/2}1_2\qquad\text{and}\qquad
\Psi_1=-x\frac{\Lambda}{x^{\dagger}x}\varphi^{-1/2},
\end{equation}
where the function $\varphi$ is fixed by the normalization condition (A.5):
\begin{equation}
\varphi =1+\frac{\Lambda^2}{x^{\dagger}x}.
\end{equation}
The relations (A.6) is verified by direct calculation. Hence, our $(\Delta,\Psi)$ satisfies
all conditions (A.2)--(A.6), and we can define a gauge potential via (A.7) and obtain from
(A.8) a self-dual gauge field on $\mathbb R^8$.
\par
Now we choose $r=k=n=2$. For the ADHM ingredients we propose the constant $8\times4$ matrices
\begin{equation}
a=\begin{pmatrix}\Lambda_0+\Lambda_1\\Q_0+Q_1\end{pmatrix},\qquad b_{i}=\begin{pmatrix}
0\\-E_{i}\end{pmatrix},
\end{equation}
where $\Lambda_{i}$ is a real matrix, $E=E_1+E_2$ is the identity matrix, and $i=0,1$. (Here
and below, we use the symbols $S_0$ and $S_1$ for the $4\times4$ matrix of the form
\begin{equation}
\begin{pmatrix} s&0\\0&0
\endpmatrix\qquad\text{and}\qquad
\pmatrix 0&0\\0&s
\end{pmatrix},
\end{equation}
where $s=s^{\mu_{i}}e^{\dagger}_{\mu_{i}}$, respectively). It is obvious that the matrix
\begin{equation}
\Delta^{\dagger}\Delta=\Lambda_{i}\Lambda_{i} +(Q_{i}-x_{i}E_{i})^{\dagger}(Q_{i}-x_{i}E_{i})
\end{equation}
is real and nondegenerate. Hence, the conditions (A.2) and (A.3) are true. In order that to
construct a solution of the equations (5) we must find a matrix $\Psi=\Psi(x)$ satisfying the
conditions (A.4)--(A.6). Suppose
\begin{equation}
\Psi=\sum^{1}_{i=0}\begin{pmatrix}-E_{i}\\U_{i}\end{pmatrix}W_{i},
\end{equation}
where $W_0$ and $W_1$ are real $4\times4$ matrices. Then by direct calculation we get that the
matrix (14) satisfies the conditions (A.4) and (A.5) if and only if the nonzero elements
$\lambda_{i}$, $q_{i}$, $u_{i}$ and $w_{i}$ of the matrices $\Lambda_{i}$, $Q_{i}$, $U_{i}$
and $W_{i}$ respectively are connected by the following relations
\begin{align}
u_{i}^{\dagger}&=\lambda_{i}(q_{i}-x_{i})^{-1},\\
w_{i}^2&=(1+u_{i}^{\dagger}u_{i})^{-1} ,
\end{align}
where we do not sum on the recurring indices and the difference $q_{i}-x_{i}\ne0$. Using (15)
and (16) we easily prove the completeness relations (A.6). Hence, our $(\Delta,\Psi)$
satisfies all conditions (A.2)--(A.6), and we can obtain from (A.8) a self-dual gauge field on
$\mathbb R^8$. Note that one may restrict our solutions to a subspace $\mathbb
R^4\subset\mathbb R^8$. In this case we get the 't Hooft-type instanton solutions in four
dimensions.
\par
Note that generalizations of the solution (9) have been described in the papers [16,17]. The
construction of a solution which generalizes (14) can be found in~[18]. However for our
purposes this will not be necessary.

\section{Heterotic string solitons}

As in the Refs.~[1]--[6] we search for a solution to lowest nontrivial order in $\alpha'$ of
the equations of motion that follow from the bosonic action
\begin{equation}
S=\frac{1}{2k^2}\int d^{10}x\,\sqrt{-g}e^{-2\phi}\left(R+4(\nabla\phi)^2
-\frac{1}{3}H^2-\frac{\alpha'}{30}\text{Tr}F^2\right),
\end{equation}
where the three-form antisymmetric field strength is related to the two-form potential by the
familiar anomaly equation
\begin{equation}
H=dB+\alpha'\left(\omega_3^{L}(\Omega)-\frac{1}{30}\omega_3^{YM}(A)\right)+\dots,
\end{equation}
where $\omega_3$ is the Chern-Simons three-form and the connection $\Omega_{M}$ is a
non-Riemannian  connection related to the usual spin connection $\omega$ by
\begin{equation}
\Omega_{M}^{AB}=\omega_{M}^{AB}-H_{M}^{AB}.
\end{equation}
We are interested in solutions that preserve at least one supersymmetry. This requires that in
ten dimensions there exist at least one Majorana-Weyl spinor $\epsilon$ such that the
supersymmetry variations of the fermionic fields vanish for such solutions
\begin{align}
\delta\chi&=F_{MN}\Gamma^{MN}\epsilon,\\
\delta\lambda&=(\Gamma^{M}\partial_{M}\phi-\frac16H_{MNP}\Gamma^{MNP})\epsilon,\\
\delta\psi_{M}&=(\partial_{M}+\frac14\Omega_{M}^{AB}\Gamma_{AB})\epsilon,
\end{align}
where $\chi$, $\lambda$ and $\psi_{M}$ are the gaugino, dilatino and gravitino fields,
respectively.
\par
Let us now show that our instanton solutions can be extended to a solitonic solution of the
heterotic string. Consider the action of the ten dimensional low energy effective theory of
the heterotic string. The bosonic part of this action is (17). If we have the solution (9),
then we can construct a fivebrane solution. Indeed, the supersymmetry variations are
determined by a positive chirality the Majorana-Weyl $SO(9,1)$ spinor $\epsilon$. Because of
the fivebrane structure, it decomposes under $SO(9,1)\supset SO(5,1)\times SO(4)$ as
\begin{equation}
16\to (4_{+},2_{+})\oplus (4_{-},2_{-}),
\end{equation}
where $\pm$ subscripts denote the chirality of the representation. Then the ansatz
\begin{align}
g_{\mu\nu}&=e^{2\phi}\delta_{\mu\nu},\\
H_{\mu\nu\lambda}&=-\epsilon_{\mu\nu\lambda}{}^{\sigma}\partial_{\sigma}\phi,
\end{align}
with the constant chiral spinor $\epsilon$ solves the supersymmetry equations with zero
background fermi fields provided the Yang-Mills gauge fields satisfies the instanton self-dual
condition (5). Substituting the explicit gauge field strength (A8) for the instanton (9) to
the anomalous Bianchi identity
\begin{equation}
dH=\alpha'\left(\text{tr} R\wedge R-\frac{1}{30}\text{Tr} F\wedge F\right),
\end{equation}
one obtains the following dilaton solution (cf.~[1]):
\begin{equation}
e^{-2\phi}=e^{-2\phi_0}+8\alpha'\frac{(x^{\dagger}x+2\Lambda^2)}{(x^{\dagger}x+\Lambda^2)^2}
+O(\alpha'{}^2).
\end{equation}
Note that the obtained string solution is not identical to [2]. Indeed, the translation
$x_{\mu_{i}}\to x_{\mu_{i}}+q_{\mu_{i}}$ introduces eight location parameters in our solution.
Four parameters localize the instanton in the subspace $\mathbb R^4\subset\mathbb R^8$. Other
four parameters restrict the choice of $\mathbb R^4$ in $\mathbb R^8$. Since the 5-brane is
transverse to $\mathbb R^4$, it follows that its selection in $M_{9,1}$ is not arbitrary. The
solution in [2] has not these restrictions.
\par
If we have the soliton solution (14), then we can construct a double-instanton string solution
analogue of (27). In this case the Majorana-Weyl $SO(9,1)$ spinor $\epsilon$ decomposes under
$SO(9,1)\supset SO(1,1)\times SO(4)\times SO(4)$ for the $M^{9,1}\to M^{1,1}\times M^{4}\times
M^{4}$ decomposition. The ansatz
\begin{equation}
\begin{aligned}
g_{\mu_{i}\nu_{i}}&=e^{2\phi}\delta_{\mu_{i}\nu_{i}},\\
H_{m_{i}n_{i}p_{i}} &=-\varepsilon_{m_{i}n_{i}p_{i}}{}^{s_{i}}\partial_{s_{i}}\phi,
\end{aligned}
\end{equation}
where $i=0$ or 1, solves the supersymmetry equations with zero background fermi fields.
Substituting the gauge field strength (A8) for the ansatz (14) to (26), we get the following
dilaton solution:
\begin{equation}
e^{-2\phi}=e^{-2\phi_0}+8\alpha'\frac{(x_{i}^2+2\lambda_{i}^2)}{(x_{i}^2
+\lambda^2_{i})^2}+O(\alpha'{}^2),\qquad i=0,1.
\end{equation}
If we restrict the solutions (27) and (29) to a subspace $\mathbb R^4\subset\mathbb R^8$, then
we recover the heterotic string solitons as derived in [2].
\par
Note also that there are different solutions with more worldsheet supersymmetry (cf.~[1] and
[3]). These symmetric solutions are characterized, from the spacetime point of view, by
$dH=0$. This condition requires, according to (26), that the curvature $R(\Omega)$ should
cancel against the instanton Yang-Mills field $F$. Both the algebraic effective action
arguments and the $(4,4)$ worldsheet supersymmetry arguments of [1] can be used in essentially
the same manner to demonstrate exactness of the string solutions.

\numberwithin{equation}{section}
\appendix
\section{Appendix}

Here, we give an extended ADHM construction of an $n$-instanton solutions for $u(2r)$-valued
gauge fields in $4k$ dimensions (see~[7]). This construction is based on a complex
$(2n+2r)\times 2r$ matrix $\Psi$ and a complex $(2n+2r)\times 2n$ matrix
\begin{equation}
\Delta=a+\sum^{k-1}_{i=0}b_{i}(x_{i}\otimes1_{n}),
\end{equation}
where $a$ and $b_{i}$ are constant $(2n+2r)\times 2n$ matrices and
$x_{i}=x^{\mu_{i}}e^{\dagger}_{\mu_{i}}$ is $2\times2$ matrices. These matrices must satisfy
the following conditions:
\begin{alignat}{2}
\Delta^{\dagger}\Delta&=f^{-1}&\qquad &\text{(invertibility)},\\
[\Delta^{\dagger}\Delta,V_{\mu}\otimes1_{n}]&=0&\qquad &\text{(reality)},\\
\Delta^{\dagger}\Psi&=0&\qquad &\text{(orthogonality)},\\
\Psi^{\dagger}\Psi&=1_{2r}&\qquad &\text{(normalization)},\\
\Psi\Psi^{\dagger}+\Delta f\Delta^{\dagger}&=1_{2n+2r}&\qquad &\text{(completeness)}.
\end{alignat}
The completeness relations (A.6) means that $\Psi\Psi^{\dagger}$ and $\Delta
f\Delta^{\dagger}$ are projectors onto orthogonal complementing subspaces of $\mathbb
C^{2n+2r}$. For $(\Delta,\Psi)$ satisfying (A.2)--(A.6) the gauge potential is chosen in the
form
\begin{equation}
A=\Psi^{\dagger}d\Psi.
\end{equation}
Indeed, after straightforward calculation the components of the gauge field $F$ then take the
form
\begin{equation}
F_{\mu\nu}=2\Psi^{\dagger}bN_{\mu\nu}fb^{\dagger}\Psi,
\end{equation}
where the $(2n+2r)\times 2nk$ matrix $b=(b_0\dots b_{k-1})$ and $\mu,\nu=0,\dots,k-1$. It is
obvious that for $k=2$ the field strength (A.8) satisfies the self-dual Yang-Mills equations
(5).

\end{document}